\documentclass[%
 reprint,
superscriptaddress,
 amsmath,amssymb,
 aps,
]{revtex4-2}

\usepackage{graphicx}
\usepackage{dcolumn}
\usepackage{bm}

\usepackage[utf8]{inputenc}
\usepackage[T1]{fontenc}

\newcommand{\liber}[1]{{\fontfamily{LibertinusSerif-TLF}\selectfont #1}}

\usepackage{lmodern}
\usepackage{etoolbox}
\usepackage{multirow}
\usepackage[normalem]{ulem}
\usepackage[table]{xcolor}
\usepackage{colortbl}

\usepackage{comment}
\usepackage{float}
\usepackage[table]{xcolor}
\usepackage{array}
\usepackage{makecell}
\usepackage{tabularx}
\newcolumntype{C}[1]{>{\centering\arraybackslash}m{#1}}
\newcolumntype{Y}{>{\centering\arraybackslash}X}
\usepackage{siunitx}
\usepackage{xcolor}
\def\wcm{{W cm$^{-2}$ }}

\makeatletter
\def\@email#1#2{%
 \endgroup
 \patchcmd{\titleblock@produce}
  {\frontmatter@RRAPformat}
  {\frontmatter@RRAPformat{\produce@RRAP{*#1\href{mailto:#2}{#2}}}\frontmatter@RRAPformat}
  {}{}
}%




\usepackage[colorlinks=true,
            linkcolor=blue,
            citecolor=blue,
            urlcolor=blue]{hyperref}

\usepackage{orcidlink}

\begin{document}

\title{Tens of MeV, collimated, bright fluxes of protons from ordered nano-structured targets in ultra-relativistic laser-matter interaction}

\author{Sagar Dam \orcidlink{0009-0009-2829-4008
}}
\thanks{Email: \textcolor{blue}{sagar.dam@tifr.res.in}}
\affiliation{Tata Institute of Fundamental Research, Colaba, Mumbai 400005, India}

\author{Stefania Ionescu \orcidlink{0009-0006-7655-8715}}
\thanks{Email: \textcolor{blue}{stefania.ionescu@eli-np.ro}}
\affiliation{ Extreme Light Infrastructure - Nuclear Physics (ELI-NP), ``Horia Hulubei'' National Institute for R\&D in Physics and Nuclear Engineering (IFIN-HH), 30 Reactorului Street, Bucharest-M\u{a}gurele, 077125, Romania}
\affiliation{National University of Science and Technology POLITEHNICA Bucharest, Splaiul Independentei no. 313, Bucharest, Romania}

\author{Jian Fuh Ong \orcidlink{0000-0003-4526-626X}}
\thanks{Email: \textcolor{blue}{jianfuh.ong@eli-np.ro}}
\affiliation{ 
Extreme Light Infrastructure - Nuclear Physics (ELI-NP), ``Horia Hulubei'' National Institute for R\&D in Physics and Nuclear Engineering (IFIN-HH), 30 Reactorului Street, Bucharest-M\u{a}gurele, 077125, Romania}

\author{Ameya Parab \orcidlink{0009-0008-5440-9991}}
\thanks{Email: \textcolor{blue}{ameya.parab@tifr.res.in}}
\affiliation{Tata Institute of Fundamental Research, Colaba, Mumbai 400005, India}

\author{Sk Rakeeb \orcidlink{0009-0006-6457-3336}}
\affiliation{Tata Institute of Fundamental Research, Colaba, Mumbai 400005, India}

\author{Hideaki Habara \orcidlink{0000-0002-7697-6830}}
\affiliation{University of Osaka, Yamadaoka, Suita, Osaka 565-0871, Japan}

\author{Gabriel Cojocaru \orcidlink{0000-0002-5877-2607}}
\affiliation{ 
Extreme Light Infrastructure - Nuclear Physics (ELI-NP), ``Horia Hulubei'' National Institute for R\&D in Physics and Nuclear Engineering (IFIN-HH), 30 Reactorului Street, Bucharest-M\u{a}gurele, 077125, Romania}
\affiliation{National Institute for Laser, Plasma and Radiation Physics, CETAL-PW Department, 409 Atomistilor Street, Magurele, Romania, 077125}

\author{Vojtěch Horný \orcidlink{0000-0002-4510-3770}}
\affiliation{ 
Extreme Light Infrastructure - Nuclear Physics (ELI-NP), ``Horia Hulubei'' National Institute for R\&D in Physics and Nuclear Engineering (IFIN-HH), 30 Reactorului Street, Bucharest-M\u{a}gurele, 077125, Romania}

\affiliation{Faculty of Nuclear Sciences and Physical Engineering, Czech Technical University in Prague, Břehová 7, 115 19 Prague 1, Czechia}

\author{Dmitrii Nistor \orcidlink{0009-0008-2193-8555}}
\affiliation{ 
Extreme Light Infrastructure - Nuclear Physics (ELI-NP), ``Horia Hulubei'' National Institute for R\&D in Physics and Nuclear Engineering (IFIN-HH), 30 Reactorului Street, Bucharest-M\u{a}gurele, 077125, Romania}
\affiliation{National University of Science and Technology POLITEHNICA Bucharest, Splaiul Independentei no. 313, Bucharest, Romania}

\author{Saidbek Norbaev \orcidlink{0009-0009-1521-5948}}
\affiliation{ 
Extreme Light Infrastructure - Nuclear Physics (ELI-NP), ``Horia Hulubei'' National Institute for R\&D in Physics and Nuclear Engineering (IFIN-HH), 30 Reactorului Street, Bucharest-M\u{a}gurele, 077125, Romania}

\author{Rudrajyoti Palit \orcidlink{0000-0001-9564-5431}}
 \affiliation{Tata Institute of Fundamental Research, Colaba, Mumbai 400005, India}

\author{Daniel Popa}
\affiliation{ 
Extreme Light Infrastructure - Nuclear Physics (ELI-NP), ``Horia Hulubei'' National Institute for R\&D in Physics and Nuclear Engineering (IFIN-HH), 30 Reactorului Street, Bucharest-M\u{a}gurele, 077125, Romania}

\author{Deepak Sangwan \orcidlink{0000-0002-7544-2028}}
\affiliation{ 
Extreme Light Infrastructure - Nuclear Physics (ELI-NP), ``Horia Hulubei'' National Institute for R\&D in Physics and Nuclear Engineering (IFIN-HH), 30 Reactorului Street, Bucharest-M\u{a}gurele, 077125, Romania}

\author{Klaus Spohr \orcidlink{0000-0001-7660-719X}}
\affiliation{ 
Extreme Light Infrastructure - Nuclear Physics (ELI-NP), ``Horia Hulubei'' National Institute for R\&D in Physics and Nuclear Engineering (IFIN-HH), 30 Reactorului Street, Bucharest-M\u{a}gurele, 077125, Romania}

\author{Bianca Stan}
\affiliation{ 
Extreme Light Infrastructure - Nuclear Physics (ELI-NP), ``Horia Hulubei'' National Institute for R\&D in Physics and Nuclear Engineering (IFIN-HH), 30 Reactorului Street, Bucharest-M\u{a}gurele, 077125, Romania}

\author{Antonia Toma \orcidlink{0009-0000-0430-3058}}
\affiliation{ 
Extreme Light Infrastructure - Nuclear Physics (ELI-NP), ``Horia Hulubei'' National Institute for R\&D in Physics and Nuclear Engineering (IFIN-HH), 30 Reactorului Street, Bucharest-M\u{a}gurele, 077125, Romania}

\author{Lucian Tudor \orcidlink{0000-0002-7762-516X}}
\affiliation{ 
Extreme Light Infrastructure - Nuclear Physics (ELI-NP), ``Horia Hulubei'' National Institute for R\&D in Physics and Nuclear Engineering (IFIN-HH), 30 Reactorului Street, Bucharest-M\u{a}gurele, 077125, Romania}

\author{Daniel Ursescu \orcidlink{0000-0002-0612-670X}}
\affiliation{ 
Extreme Light Infrastructure - Nuclear Physics (ELI-NP), ``Horia Hulubei'' National Institute for R\&D in Physics and Nuclear Engineering (IFIN-HH), 30 Reactorului Street, Bucharest-M\u{a}gurele, 077125, Romania}

\author{Adrian Vatcu}
\affiliation{ 
Extreme Light Infrastructure - Nuclear Physics (ELI-NP), ``Horia Hulubei'' National Institute for R\&D in Physics and Nuclear Engineering (IFIN-HH), 30 Reactorului Street, Bucharest-M\u{a}gurele, 077125, Romania}

\author{Keita Yamanaka \orcidlink{0009-0003-0441-3224}}
\affiliation{University of Osaka, Yamadaoka, Suita, Osaka 565-0871, Japan}

\author{Prashant Kumar Singh \orcidlink{0000-0003-3748-5411}}
\affiliation{Tata Institute of Fundamental Research Hyderabad, 36/P, Gopanpally Village, Serilingampally Mandal, Hyderabad, Telangana 500046, India}

\author{Kazuo. A. Tanaka \orcidlink{0000-0002-1493-2509}}
\affiliation{ 
Extreme Light Infrastructure - Nuclear Physics (ELI-NP), ``Horia Hulubei'' National Institute for R\&D in Physics and Nuclear Engineering (IFIN-HH), 30 Reactorului Street, Bucharest-M\u{a}gurele, 077125, Romania}
\affiliation{University of Osaka, Yamadaoka, Suita, Osaka 565-0871, Japan}

\author{G. Ravindra Kumar \orcidlink{0000-0002-2324-2825
}}
    \thanks{Email: \textcolor{blue}{grk@tifr.res.in}}
    \thanks{(corresponding author)}
    \affiliation{Tata Institute of Fundamental Research, Colaba, Mumbai 400005, India}

\date{\today}

\begin{abstract}
Laser-driven proton acceleration from nanostructured solid targets has been extensively studied, yet its performance under realistic temporal contrast conditions at petawatt-class facilities remains an open question. We present an experimental investigation of proton generation from nanostructured and flat solid targets performed at the ELI-NP facility using femtosecond laser pulses at peak intensities of $\sim 3\times10^{21}$ \wcm. Proton spectra are compared for two contrast regimes: $\sim 10^{-10}$ without plasma mirror and $\sim 10^{-13}$ with single plasma mirror. Importantly, measurable enhancement in the cutoff energy persists for the nanowire targets at both contrast levels, indicating robustness of nanowire targets against moderate pre-pulse intensities. Alongside, study of energy resolved angular distribution reveals that nanowires promote more directional emission with higher flux of high-energy protons along the target normal, while flat targets produce broader angular distributions. The results are well supported and explained by 3D particle-in-cell simulations.\\[4pt]

\end{abstract}

\maketitle

\section{Introduction}

\begin{figure*}
\centering
\includegraphics[width=\textwidth]{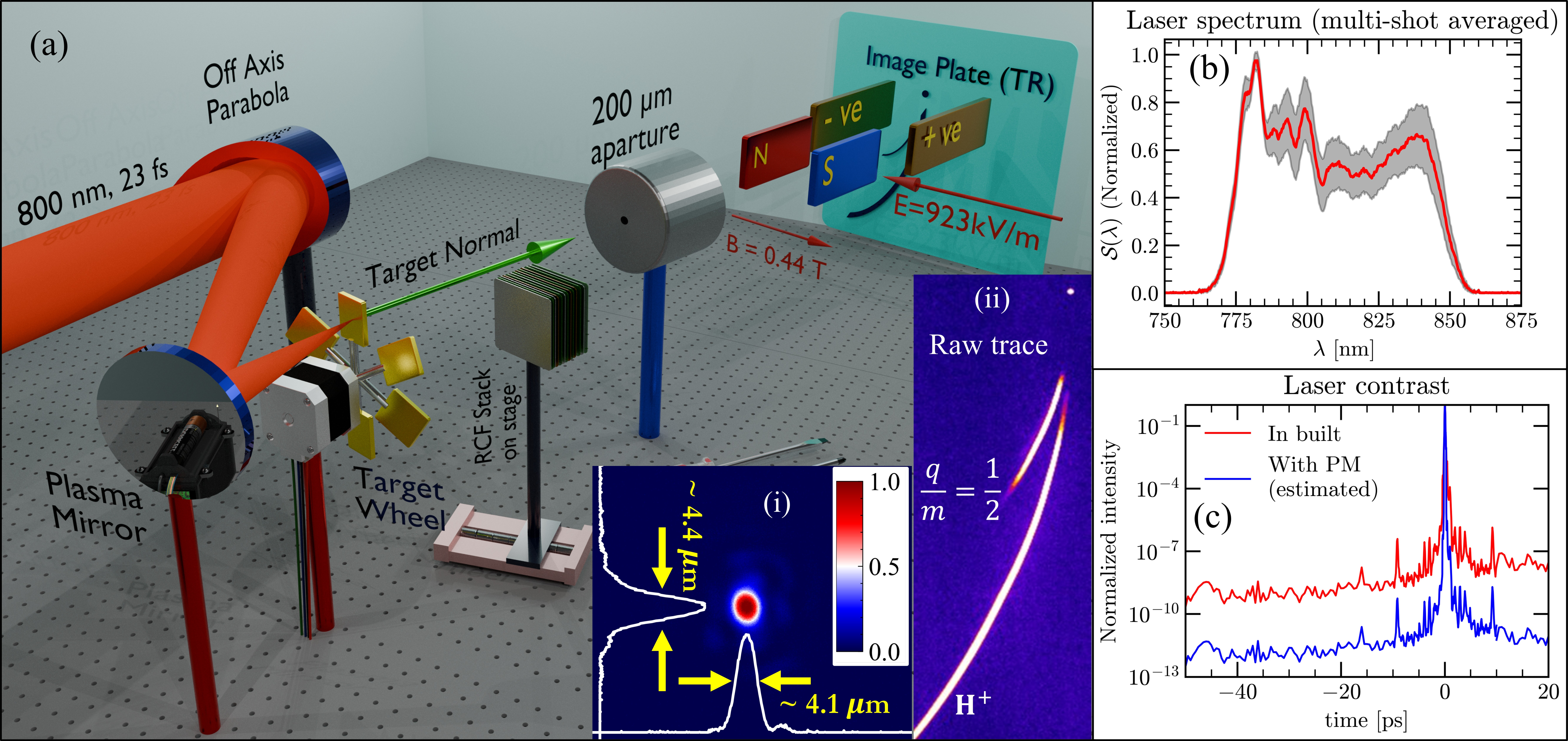}
\caption{\liber{The experimental setup is shown in (a). The on-target energy is around 12 J. The RCF stack is moved in and out from target back when required. In the insets: the focal spot image (i) taken by an imaging camera gives a spatial full width at half maximum (FWHM) of $\sim 4.1-4.4\ \mu$m. The colorbar represents normalized intensity. And (ii) the raw TP trace from the experiment. The laser spectrum (b) has a bandwidth of around 70 nm (multi shot averaged). The intensity contrast plot (c) shows the pulse has a temporal contrast of $\sim 10^{-10}$ at 50 ps time. The estimated temporal contrast with the plasma mirror at the same time is $\sim10^{-13}$.}}
\label{setup}
\end{figure*}

Ultra-intense laser-solid interactions  \cite{fortov_Extreme_states_of_matter,gibbonbook,Kruer,eliezer2002interaction,snavely2000intense,proton_accelerator_robson2007scaling,electron_accelerator_kurz2021demonstration,electron_accelerator_hogan2005multi,electron_accelerator_litos2014high,choudhary2025generation,ICF_betti2016inertial,mulser2010high,liu2019high,ICF_myatt2014multiple,chatterjee2017micron,ankit2022spectralinterferometry,rocca2024ultra,Dam2026PRR} at petawatt-class facilities have, over the past two decades, established laser-driven proton acceleration  \cite{schreiber2006analytical,borghesi2008laser,gitomer1986fast,fews1994plasma,maksimchuk2000forward,beg1997study,clark2000energetic,clark2000measurements,snavely2000intense,ion_acceleration_macci_RMP_review,proton_accelerator_robson2007scaling,badziak2018laser,kluge2012high,daido2012review,willingale2007ion,romagnani2005dynamics,mulser2010high} with proton energies reaching as high as 150 MeV  \cite{Ziegler2024} as a mature yet rapidly evolving field, with demonstrated relevance to high-energy density physics, nuclear physics, and applied particle sources. A substantial amount of work on both theoretical and experimental ground has explored the use of nanostructured targets  \cite{cowan2004ultralow_TNSA,passoni2016toward,margarone2015laser,floquet2013micro,margarone2012laser,fedeli2018ultra,andreev2011efficient,klimo2011short,yu2012laser,lubcke2017prospects,ji2017exploring,passoni2019advanced,bagchi2007fast,giuffrida2017manipulation,torrisi2016nanostructured,bagchi2012surface,cristoforetti2017transition,dalui2015preferential,vallieres2019enhanced,ebert2017laser,vallieres2021enhanced,blanco2017table,dozieres2019optimization} to improve laser energy absorption and proton yields. However, most of these studies have been performed under exceptionally high temporal contrast conditions  \cite{mourou_Tajima_2006optics_RMP,kiriyama2012temporal,choi2020highly}, often well beyond those routinely available in emerging multi-PW facilities. As a result, the practical robustness of nanostructured targets under realistic operating contrast levels and the dynamics during the interaction with the peak of the pulse still remains insufficiently understood.

In this context, the present study reports a systematic proton-acceleration experiment on solid nanostructured targets conducted at ELI-NP \cite{tanaka2020current}, under the facility’s prevailing contrast conditions and experimental constraints. Our study pertains to the widely investigated target normal sheath acceleration (TNSA) regime  \cite{wilks2001energetic_TNSA}. By directly comparing two experimentally relevant contrast regimes - approximately $10^{-10}$ without a plasma mirror (PM) and $10^{-13}$ with a plasma mirror \cite{inoue2016single,choi2020highly,contrast1998itatani,plasmamirror2006wittmanntowards,plasmamirror2018foldes}, this work provides specific insight into how target nanostructuring performs under conditions that are representative of present and near-future operation at ELI-NP and similar multi-PW laser systems and the possibilities of even higher contrast conditions.

Beyond a straightforward contrast comparison, the study addresses several open questions of broad relevance. First, it demonstrates that nanostructure-induced proton enhancement is not significantly affected at moderate contrast ($\sim 10^{-10}$), even at peak intensities approaching $3\times 10^{21}$ \wcm, calming the concern that nanostructures become ineffective unless extreme contrast cleaning is employed \cite{rocca2024ultra}. This finding establishes an important degree of robustness in nanowire targets, indicating tolerance to pre-pulse intensities that may locally reach $10^{12}-10^{13}$ \wcm due to field enhancement \cite{rajeev2004nanostructures} at sharp features. Second, by comparing energy dependent angular distributions, the work shows how the nanostructures guide the higher energy protons and give a much more enhanced and directional emission of particle beams. Third, the study compares the results of {nine} different kinds of structured and non-structured targets and gives an overview of the interaction and effectiveness of the response of target geometry.

The inclusion of particle-in-cell simulations further strengthens the ground of these experimentally observed results. The simulation closely reproduces the enhancement of both the proton yield (at the higher energy part of the spectrum) and the energy cutoff. Also the simulation shows how the enhanced azimuthal magnetic field at the back of the target guides the particles to a much narrower emission cone. These observations have direct implications for the detector placement, beam transport and applications requiring directional proton sources.

Taken together, this study serves not merely as a  demonstration of the feasibility of nanostructures for ion acceleration, but also as a validation of operating space for petawatt-class laser-solid experiments at ELI-NP. It establishes confidence that nanostructured targets can be deployed safely and effectively at the 1-PW level and provides an encouraging outlook for scaling up-to the forthcoming 10-PW regime, where higher proton energies, improved directionality are anticipated. In doing so, the work offers both immediate experimental guidance and a forward-looking framework for exploiting nanostructured targets at ultra-intense laser facilities.

\section{Methods}
\subsection{Experimental Details}
The experiment has been carried out with the 1 PW Ti-Sapphire laser facility at ELI-NP, Romania \cite{cernaianu2025commissioning,ELINP_Laser_2020_High_Power_Laser_Science_and_Engineering}. The laser has an average output energy of 29.5 J before the compression stage with central wavelength at 810 nm with an average pulse duration of 23 fs (measured with Frequency Resolved Optical Gating  \cite{trebino1993using,trebino1997measuring}) and a  \textcolor{black}{contrast level of $10^{-10}$} [see Fig. \ref{setup}(c)] at 50 ps, with a couple of pre-pulses around 10 ps.

Inside the experimental chamber [Fig. \ref{setup}(a)] a dielectric coated parabolic mirror with high precision control actuators was used to focus the beam with a measured diameter of 4.1-4.4 $\mu$m [Fig. \ref{setup}(a.i)]. A plasma mirror was used before the focus to enhance the contrast to a level of $10^{-13}$. After the plasma mirror the energy on target was around 12 J. The optimized focal spot was imaged with an imaging microscope. After the beam transport and plasma mirror we got an on-target average peak intensity of around $3\times10^{21}$ \wcm. The equivalent normalized vector potential  was $a_0=0.85\sqrt{I_L[10^{18}\text{ W cm}^{-2}]} \lambda_L[\mu\text{m}] \sim 36$, which is ultra-relativistic ($a_0\gg1$) interaction. The targets were mounted on a target wheel and placed at in focus with the help of an imaging microscope. We repeated this process every time we move the target wheel to use a new target. The angle of incidence on the target was maintained at $15^\circ$.

To get the ion spectrum, we used the standard Thomson parabola (TP)  method \cite{thomson1911xxvi,slater1978thomson,sakabe1980modified,treffert2018design,jung2011novel,MK_2017_RSI_tata} along the target normal direction (See supplementary \cite{SuppM} section $\S1$ for details). As per existing understanding, at the rear side, most of the ions are accelerated along the target normal direction by TNSA mechanism  \cite{wilks2001energetic_TNSA,cialfi_tnsa_phd,ion_acceleration_macci_RMP_review,hegelich2002mev_TNSA,cowan2004ultralow_TNSA,borghesi2006fast_TNSA,hegelich2006laser_TNSA,roth2013ion_TNSA,passoni2013advances,schreiber2006analytical,zeil2010scaling}. To get a collimated narrow ion beam into the TP, we used a 200 $\mu$m diameter pinhole collimator at a distance of 450 mm from the interaction spot. A pair of Nd-magnets with a length of 50 mm produces a transverse magnetic field of 0.44 T. A parallel plate copper capacitor with a plate separation of 13 mm and potential difference of 12 kV was used to produce an electric field of 923 kV/m. Finally an image plate (Fujifilm BAS-TR) \cite{IP_detector_RSI_2024} was placed after the capacitor to collect the ion trace. To protect the signal from scattered light, the image plate (IP) was covered with $10\ \mu$m aluminium (Al) foil (this will also block protons up to energy $\sim 1\ $MeV). The image plate (IP) was imaged with a Fujifilm FLA-7000 IP reader to retrieve the trace \cite{tanaka2005IPcalibration}. The time for each shot was noted carefully to retrieve the ion spectrum as the IP signal gradually fades away. The IP calibration has been taken from the previous work done by Martin \textit{et al}.  \cite{Calibration_of_TR_IP_for_ions}

Along with the Thomson parabola measurements, radiochromic film \cite{garcia2025radiochromic,diaz2024investigation} (RCF) stacks (HD-V2 + EBT-3) were employed to obtain energy dependent angularly resolved information of the accelerated proton beam. The RCF stack was mounted on a motorized stage at 2.5 cm behind the target to capture the particle beam. To eliminate noise from laser light, the stack was covered with $10\ \mu$m Al foil. The full information of the RCF stack and the calibrated energy chart is given in the supplementary  \cite{SuppM}  section $\S3$.

\noindent
\makebox[\linewidth][l]{\rule{0.4\linewidth}{0.5pt}}

\noindent
\hypertarget{Supplementary}{%
\textcolor{violet}{\textit{More details on experimental methods in supplementary material.} \cite{SuppM}}}

\subsection{Nanostructure growth}

\begin{figure}
\centering
\includegraphics[width=\linewidth]{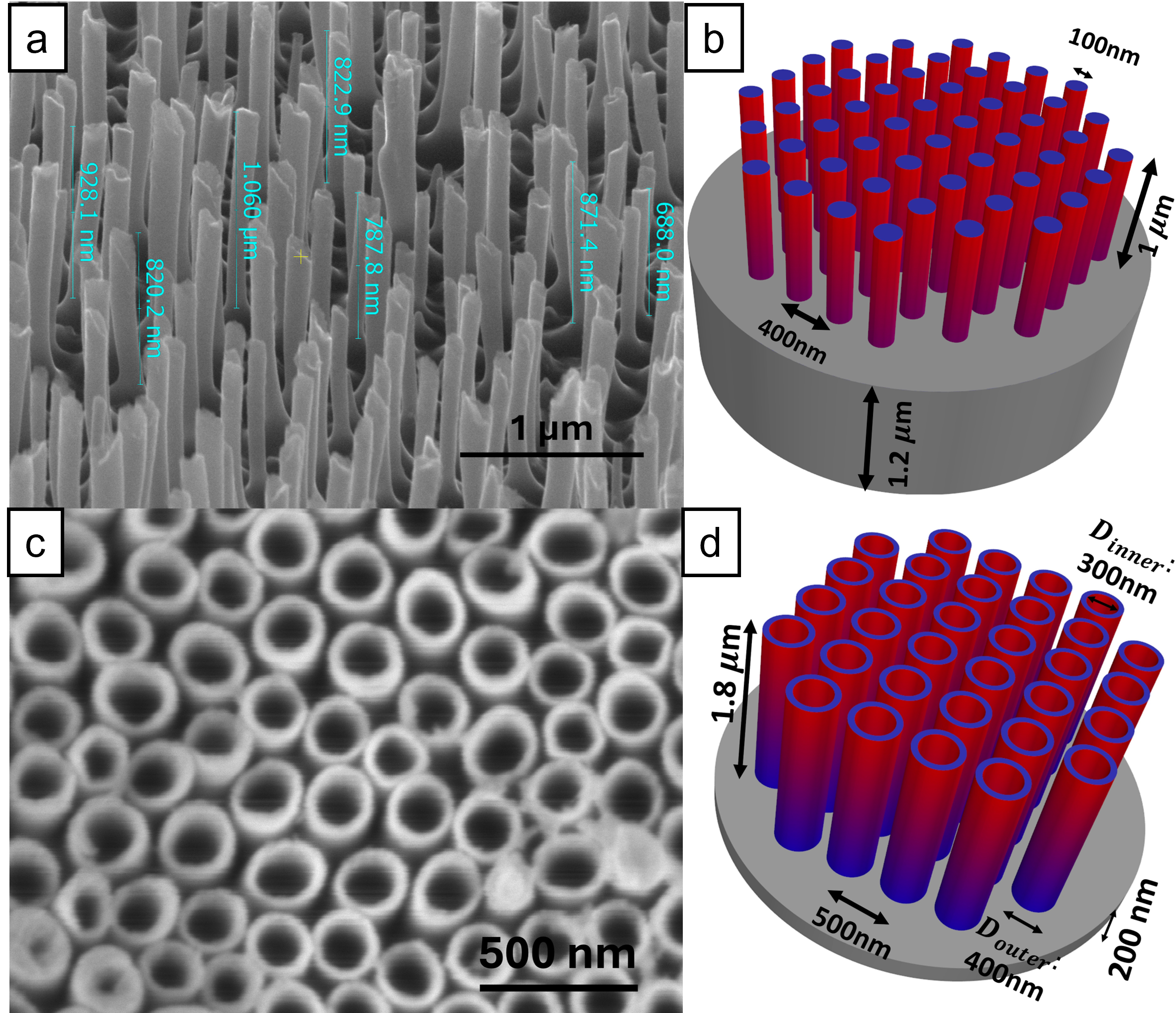}
  \caption{\label{Fig:target}
  \liber{Scanning Electron Microscope (\textsc{SEM}) images of nickel nanowires (a) and nanotubes (c) and target schemes (b,d).}
  }
\end{figure}

Metallic nanowires and nanotubes were grown by electrochemical methods [see Fig. \ref{Fig:target}(a)-(d)],  using a porous alumina template, which was obtained by aluminum anodization  in acidic electrolytes (oxalic, phosphoric and citric acid)  \cite{Gheorghiu2021}. By changing the anodization process parameters, the diameters of the pores in the alumina template as well as the gaps between the pores can be adjusted, resulting in a highly ordered honeycomb-like structure  \cite{Masuda1995}. The nanostructures are deposited in the porous alumina by electrodeposition from a solution containing the salts of the metals to be deposited. For our case, with nickel nanostructures, nickel sulfate and nickel chloride were used. The template determines the final nanostructure parameters (gap, diameter and morphology), while the deposition duration sets the length of the structures. After the deposition, the template is dissolved and the structures liberated, followed by the multi-step drying procedure in several solvents to avoid disruption of the structures and the cluster formation resulting from the surface tension of the liquids  \cite{Ionescu2025}. The resulting nickel nanowires and nanotubes are vertically aligned, highly ordered, on thin nickel substrates with thicknesses of few micrometers down to \qty{200}{\nano\meter}. The structures diameters varies between \qty{100}{\nano\meter} to \qty{400}{\nano\meter}, and the gap between the wires from \qty{250}{\nano\meter} to \qty{450}{\nano\meter}. The nanowires have a filled, solid center, while the nanotubes are hollow inside, and therefore, a reduced overall density. Furthermore, using the same electrodeposition method, flat nickel foils were prepared as a comparison target. The method allows to grow structures on large areas homogeneously, covering few $\si{\centi\meter}^2$, which after are cut to few \si{\milli\meter} pieces and mounted on the target holder.

\section{Experimental Results}

\begin{figure*}
\centering
\includegraphics[width=\textwidth]{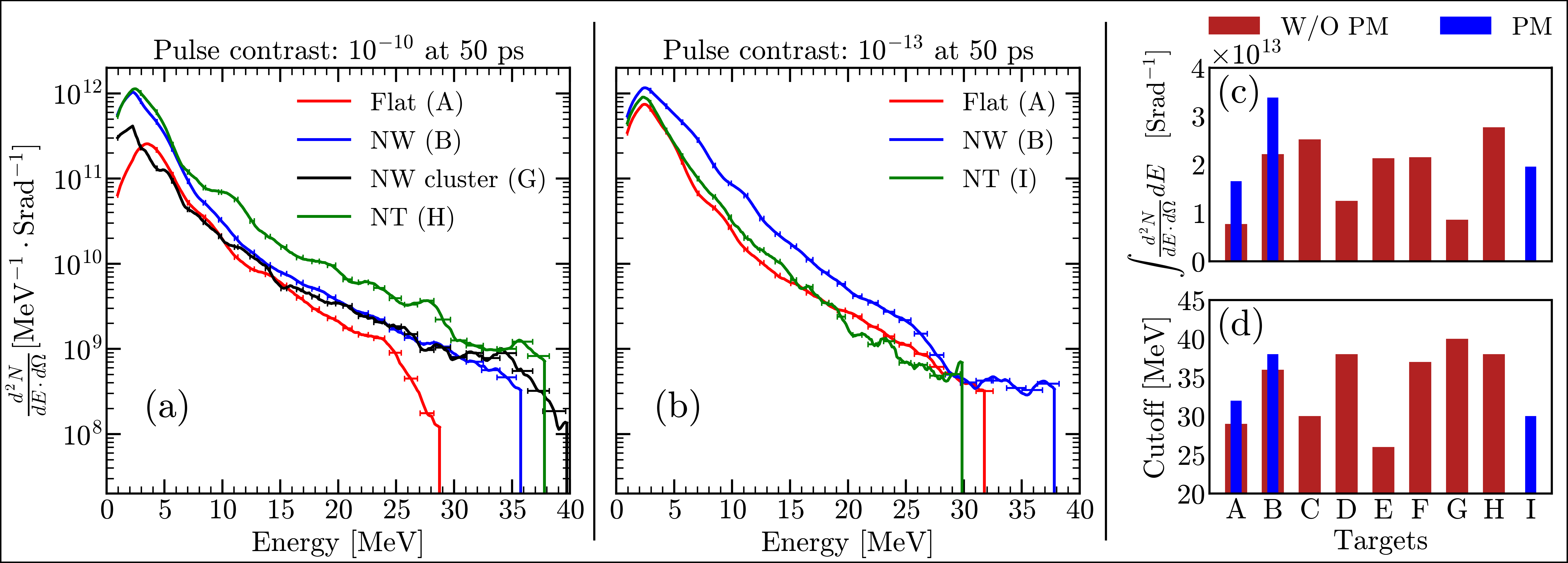}
\caption{\liber{Experimentally obtained proton energy spectra (for different flat and structured targets) along the TNSA direction using Thomson parabola method. The two temporal contrast conditions are shown as: (a) without plasma mirror ($10^{-10}$ relative intensity at 50 ps) and (b) with plasma mirror ($10^{-13}$ relative intensity at 50 ps). In the plot legends (and elsewhere mentioned in the manuscript) NW and NT represents  nanowire and nanotube targets respectively. The dimensions of different targets (A,B,C, \ldots etc) are given in table \ref{tab:target-specs}. Top right: (c) comparison of integrated proton count for different  targets. Bottom right: (d) comparison of proton cutoff energy for those targets.  With 50 ps relative intensity contrasts: red$\to 10^{-10}$ \& blue $\to10^{-13}$ (without \& with PM).}} 
\label{experimental_plots}
\end{figure*}

\subsection{Proton spectrum from Thomson parabola (TP) method}

Figure \ref{experimental_plots} (a)-(b) shows multi-shot averaged proton energy spectra obtained using the TP spectrometer for several target configurations under two different pulse contrast conditions. The target nomenclature (A,B,C, \ldots etc), geometrical parameters [substrate thickness ($T_\text{sub}$), nanostructure length and diameter ($L_\text{nano},D_\text{nano}$) gap between structures] of all the investigated targets and corresponding proton cutoff energies $\left(E^\text{expt}_\text{max}\right)$ are summarized in Table \ref{tab:target-specs}. The horizontal error bars along the energy axis represent the uncertainty due to the finite width of the TP entrance pinhole ($200\ \mu$m). The maximum proton energy cutoff is determined from the TP traces at the point where the signal at the high-energy end becomes comparable to the background noise level.

\begin{table}[h]
\centering
\footnotesize
\setlength{\tabcolsep}{4pt}
\renewcommand{\arraystretch}{1.5}

\resizebox{\columnwidth}{!}{

\begin{tabular}{|c|c|c|c|c|c|c|c|}
\hline
\hline
\textcolor{violet}{\makecell{Target\\Name}} &
\textcolor{violet}{\makecell{Target\\Type}} &
\textcolor{violet}{\makecell{$T_\text{sub}$\\($\mu$m)}} &
\textcolor{violet}{\makecell{$L_\text{nano}$\\($\mu$m)}} &
\textcolor{violet}{\makecell{$D_\text{nano}$\\($\mu$m)}} &
\textcolor{violet}{\makecell{Gap\\($\mu$m)}} &
\textcolor{violet}{\makecell{Plasma\\Mirror}} &
\textcolor{violet}{\makecell{$E_{\max}^{\text{expt}}$\\(MeV)}} \\

\hline

\multirow{2}{*}{A} & \multirow{2}{*}{Ni flat} & \multirow{2}{*}{2.0} &
\multirow{2}{*}{--} & \multirow{2}{*}{--} & \multirow{2}{*}{--}
& No & 29 \\
\cline{7-8}
& & & & & & Yes & 32 \\

\hline

\multirow{2}{*}{B} & \multirow{2}{*}{Ni NW} & \multirow{2}{*}{1.2} &
\multirow{2}{*}{1.0} & \multirow{2}{*}{0.10} & \multirow{2}{*}{0.40}
& No & 36 \\
\cline{7-8}
& & & & & & Yes & 38 \\

\hline

C & Ni NW & 0.2 & 1.3 & 0.15 & 0.45 & No & 30 \\
\hline
D & Ni NW & 1.4 & 3.8 & 0.20 & 0.45 & No & 38 \\
\hline
E & Ni NW & 3.5 & 9.5 & 0.20 & 0.40 & No & 26 \\
\hline
F & Ni NW & 0.6 & 1.6 & 0.10 & 0.40 & No & 37 \\
\hline
G & \makecell{Ni NW\\cluster$^\text{\textcolor{blue}{\#}}$} & 0.5 & 4.0 & 0.25 & 1.25 & No & 40 \\
\hline

H & Ni NT & 0.2 & 1.8 &
\makecell{$D_\text{ou}:0.2$\\ $D_\text{in}:0.1$} &
0.25 & No & 38 \\
\hline

I & Ni NT & 0.2 & 1.6 &
\makecell{$D_\text{ou}:0.4$\\ $D_\text{in}:0.3$} &
0.45 & Yes & 30 \\
\hline
\hline
\end{tabular}
}
\caption{\liber{Specifications of targets used in the experiment. 
$T_\text{sub}, L_\text{nano}, D_\text{nano}, \text{Gap}, E_\text{max}$ represent substrate thickness, nanopillar length, diameter, spacing, and maximum proton energy. 
\textcolor{black}{\textit{(NW: Nanowires; NT: Nanotubes)}}}\\
$^\text{\textcolor{blue}{\#}}$\liber{\textcolor{violet}{{Ni NW cluster: several  nanowires agglomerated in clusters/ bundles with a total diameter of 350 nm}}}}
\label{tab:target-specs}

\end{table}

Figure \ref{experimental_plots} (a) shows the proton energy spectra for flat targets (A), Ni nanowire (B), clustered Ni nanowire (G), and nanotube (H) targets with contrast of $10^{-10}$ at 50 ps. In particular, the proton cutoff energy increases from approximately 29 MeV for the flat target to about 36 MeV for the Ni nanowire target. Using the empirical TNSA scaling proposed by Zimmer \textit{et al.} \cite{ZimmerTNSAscaling}, the proton cut-off energy is expected to increase from 29.0 MeV to 31.5 MeV when the initial 2 $\mu$m-thick foil is replaced by a thinner 1.2 $\mu$m-thick foil. Therefore, the additional increase from 31.5 MeV to 36 MeV can be attributed to the presence of the nanowires. An even higher cutoff energy of approximately 40 MeV is obtained from the clustered Ni nanowire target (G). This indicates that nanostructuring of the target surface can significantly enhance the proton acceleration process through improved laser energy absorption and more efficient hot-electron generation that strengthens the sheath field (observed in the 3D PIC simulation) responsible for target-normal sheath acceleration.

The results obtained with an improved estimated contrast of $10^{-13}$ at 50 ps, using a plasma mirror are presented in Fig. \ref{experimental_plots} (b) for the same flat and nanowire targets, and nanotube target (I). Under enhanced contrast conditions, the cutoff energies for both the flat and nanowire targets do not show a noticeable increase relative to the case without a plasma mirror. This observation suggests that the enhancement associated with nanostructured targets is already present even at the lower contrast level shown in Fig. \ref{experimental_plots} (a), indicating that the beneficial effects of the nanostructure are largely preserved under moderate pre-pulse conditions. Ni nanotube targets show different behaviors in case of these two different contrast levels. NT (H) with $10^{-10}$ contrast has similar cut-off with nanowire target, while NT (I), which has bigger tube diameter than NT (H), at high contrast has a low performance, with a cut-off at 30 MeV. This can be because of a lower material density which leads to fewer electrons, and in the case of low contrast, a pre-plasma is formed and helps coupling with the main pulse, while in the high contrast, less pre-plasma is formed.

In addition to the increase in cutoff energy, structured targets also lead to a substantial enhancement in the overall proton yield. This effect is illustrated more clearly in Fig. \ref{experimental_plots} (c), which shows the energy-integrated proton flux for all investigated targets. It can be seen that all structured targets produce a significantly larger number of protons compared to the flat target under both contrast conditions. Figure \ref{experimental_plots} (d) summarizes the cutoff energies obtained for all targets for easier comparison. The lowest cutoff energy is observed for target E, corresponding to the nanowire target with the largest substrate thickness. This behavior is consistent with previous studies showing that the maximum proton energy in target-normal sheath acceleration decreases with increasing target thickness, due to the reduction of the accelerating sheath field and increased electron transport losses  \cite{ion_acceleration_macci_RMP_review}. Together, these observations demonstrate that nanostructured targets not only enhance the maximum proton energy but also significantly increase the proton flux, while maintaining their performance even under moderate laser contrast conditions.

\subsection{RCF traces}

\begin{figure}[h]
    \centering
    \includegraphics[width=\linewidth]{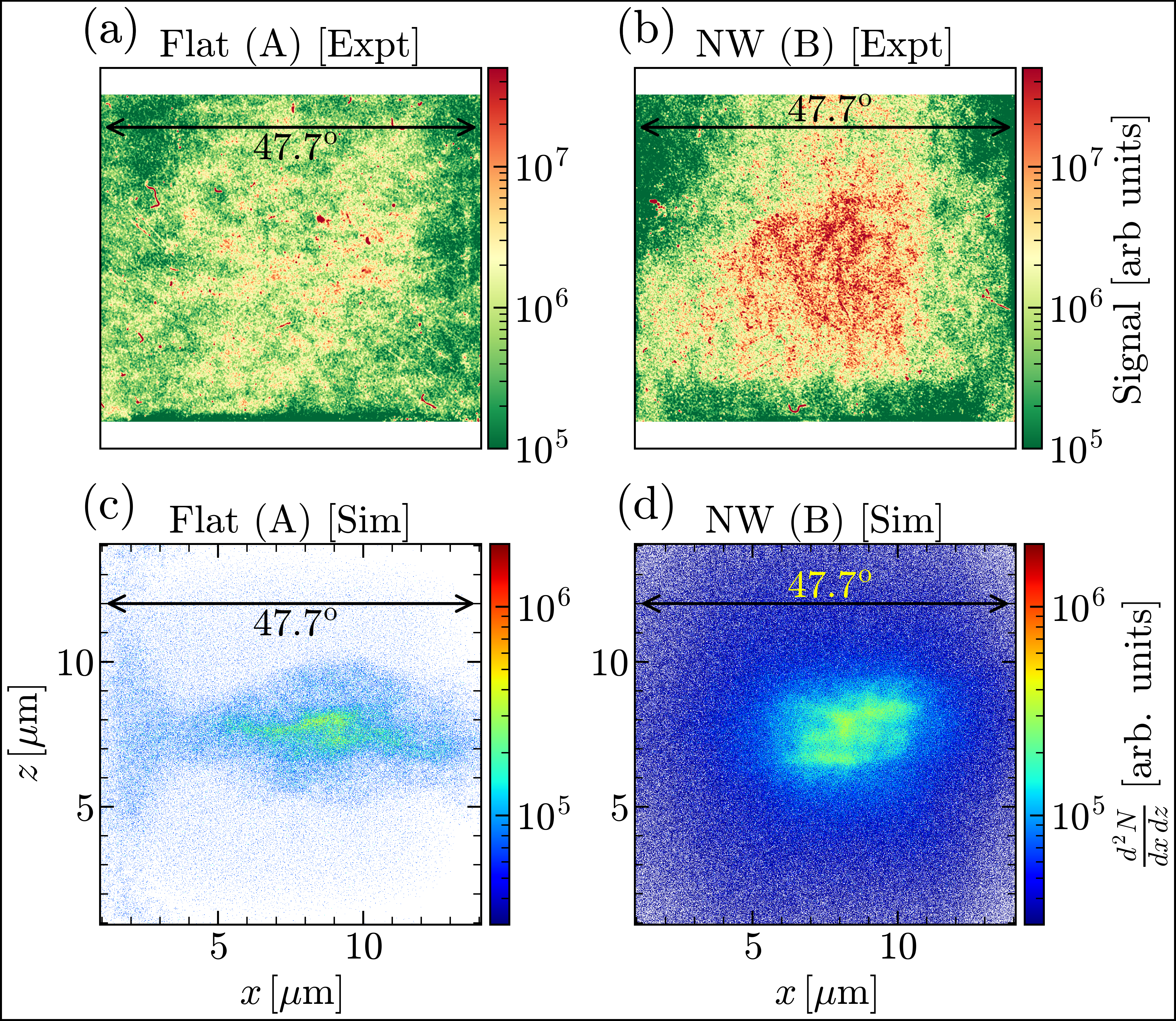}
    \caption{\liber{Experimentally obtained RCF traces capturing proton with $E>12$ MeV \iffalse with contrast $10^{-13}$ at 50 ps (with PM condition)\fi for (a) flat target (target-A) and (b) nanowire target (target-B). The directionality and flux enhancement is clearly visible for nanowire target. In the bottom panel -  (c),(d) shows the simulated result (see \ref{sec:simulation results} for details) for spatial distribution of protons in target normal plane (at a   distance of 14.7 $\mu$m from target) with the proton count integrated in the target normal forward direction. \iffalse at time of \textcolor{red}{?? ps}\fi}}
    \label{fig:raw RCF}
\end{figure}

\begin{figure}[h]
    \centering \includegraphics[width=0.98\linewidth]{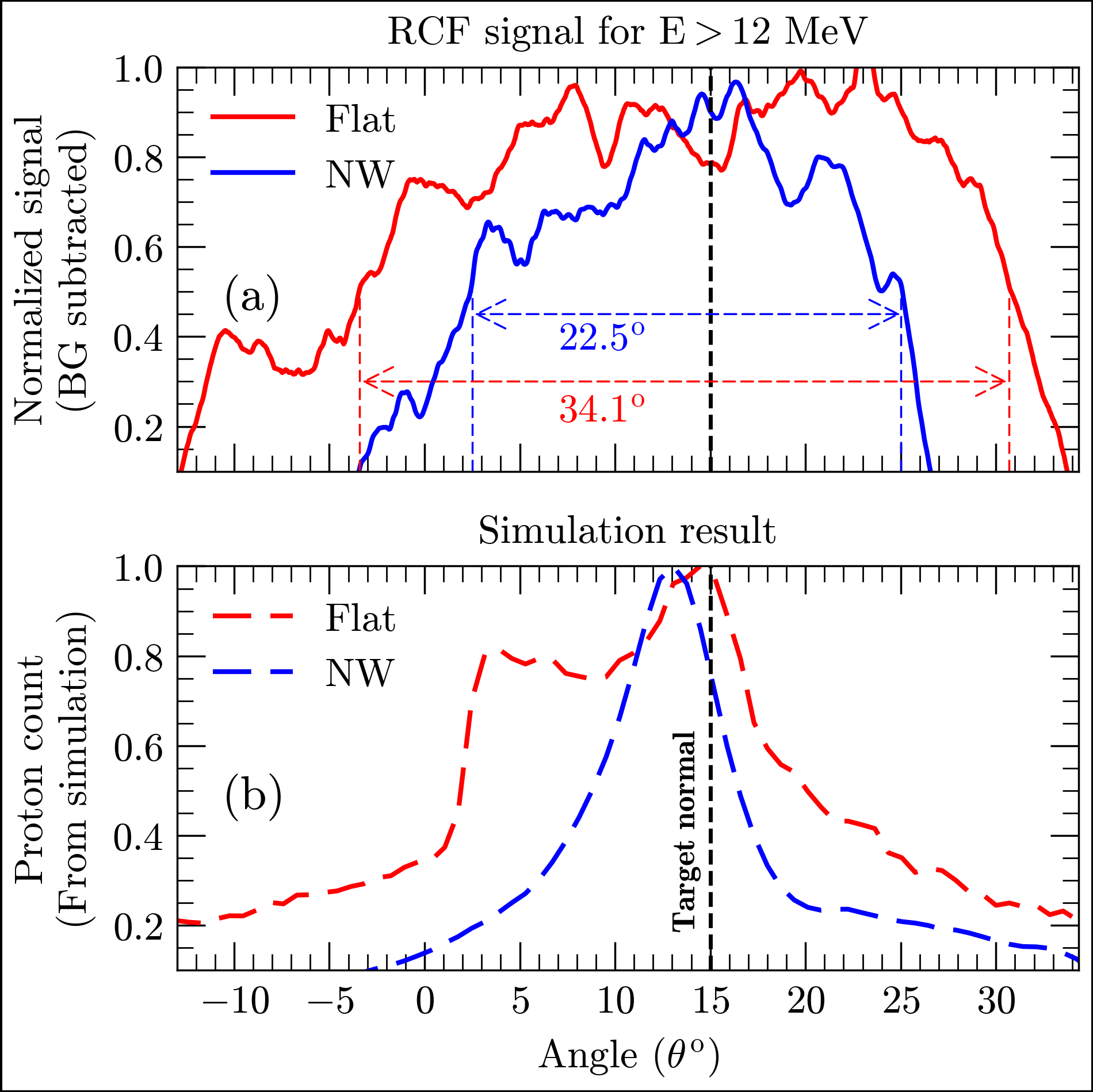}
    \caption{\liber{The top panel (a) represents the comparison of angular distribution of protons (with energy $>12$ MeV) for flat (A) and nanostructured (B) targets from experimentally obtained RCF signal. The FWHM of the angular distributions have been indicated by the corresponding colored arrows and written in the plot. For comparison the bottom panel (b) represents the simulated results for the same. The laser propagation direction is taken as $0^\circ$ angle. For both plots, y axis is in normalized unit.}}
    \label{Fig:RCF image}
\end{figure}

Figures \ref{fig:raw RCF} (a) and (b) present artificially color-mapped and inverted RCF images acquired for flat target (A) and nanowire target (B), respectively, for proton energies exceeding 12~MeV. A pronounced enhancement of the proton signal is observed for the nanowire target relative to the flat target (as indicated by the colorbar scale), in agreement with the increased proton flux measured by the TP spectrometer for proton energies above 12~MeV.

In addition, the RCF diagnostics reveal a noticeable difference in the angular characteristics of the proton emission. As shown in figure \ref{Fig:RCF image} (a), protons from the nanowire target at this energy exhibit a more directional angular distribution compared to those from the flat target. This is quantitatively very clear as the full width half maxima of the angular divergence is significantly wider for flat target ($34.1^\circ$) while compared to the nanowired target ($22.5^\circ$). This observation is qualitatively in agreement with the particle-in-cell simulation results presented in figure ~\ref{Fig:RCF image} (b), which predict enhanced directionality of high-energy protons for nanostructured targets due to the formation of stronger and more localized accelerating sheath fields. The detail of simulations are presented in the following section.

\section{simulation results\label{sec:simulation_results}}
\begin{figure*}
\centering
\includegraphics[width=\textwidth]{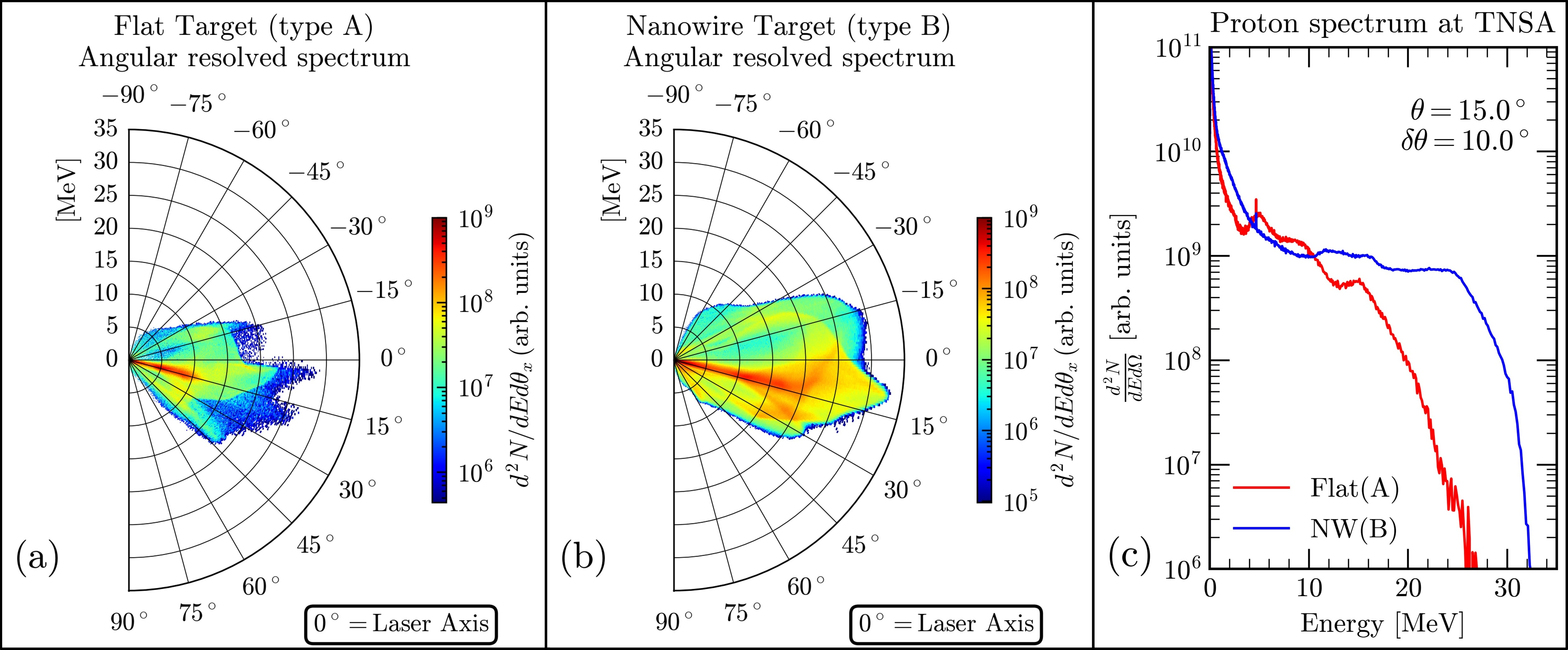}
\caption{\liber{Results of 3D PIC simulations. The angular resolved proton energy spectra of proton for (a) flat [target A] and (b) nanowire [target B] targets. The laser propagation direction is at $0^\circ$. (c) The proton energy spectra for flat (red solid line) and nanowire (blue solid line) at target normal direction integrated within $\pm 10^\circ$ cone.}}
\label{simulation_plots}
\end{figure*}

\begin{figure*}
    \centering
\includegraphics[width=\textwidth]{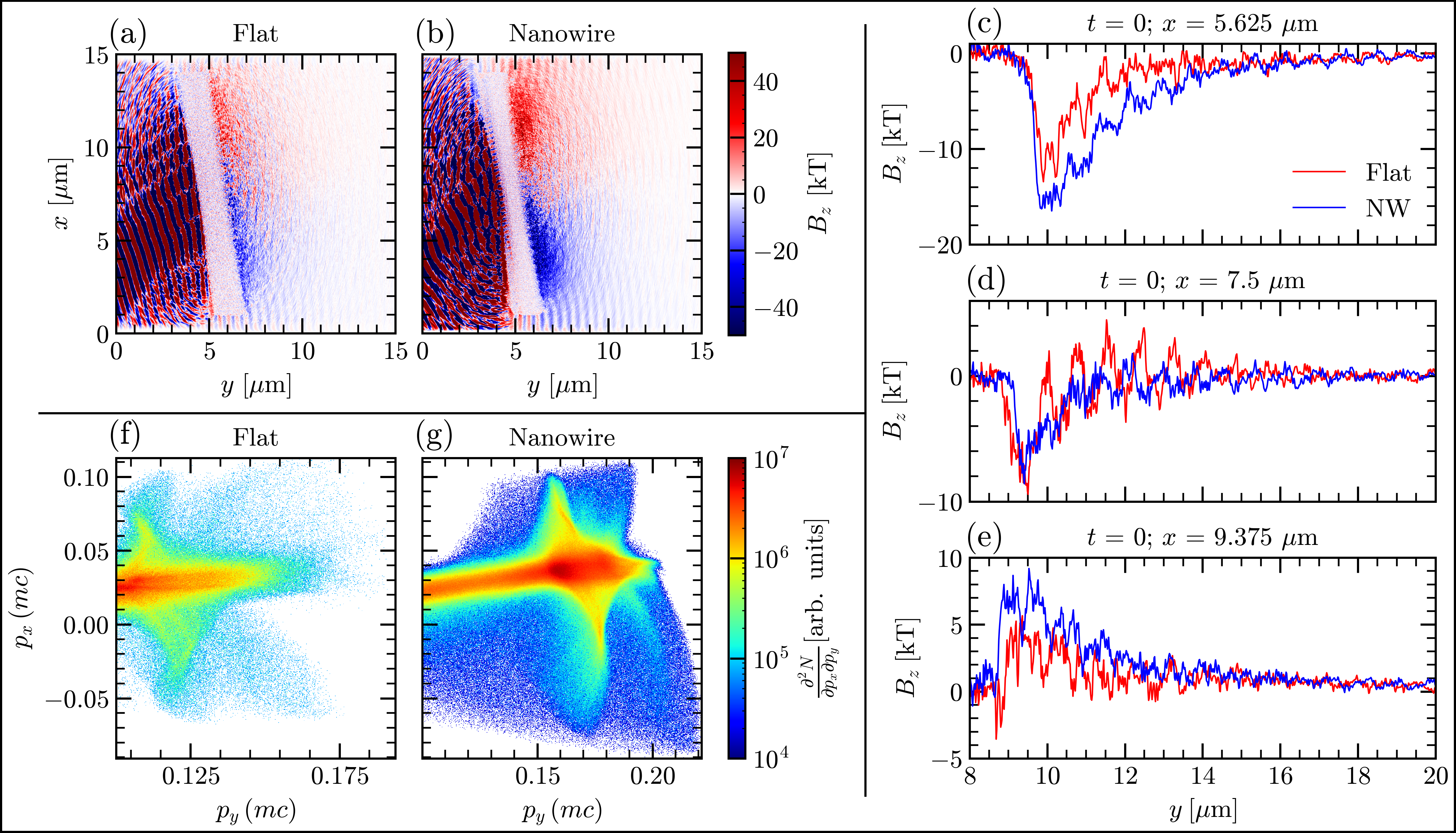}
\caption{\liber{Top left: 2D central slice of magnetic field, $B_z$ in the $y-x$ ($z-$axis points into the page) plane for (a) flat, and (b) nanowire targets. The magnetic field at the target rear shows clear enhancement in nanowire target. Right: Lineout of the of the magnetic field, $B_z$ along the $y-$axis at (c) $x=5.265\, \mathrm{\mu m}$, (d) $x=7.5\, \mathrm{\mu m}$, and (e) $x=9.375\, \mathrm{\mu m}$ at $t=0$. Bottom left: Proton momentum phase-space, $p_y-p_x$ for (f) flat, and (g) nanowire targets.}}
\label{fig:B_field}
\end{figure*}

To model the behavior of proton acceleration, coupled three-dimensional radiation-hydrodynamic (3D RHD) and particle-in-cell (3D PIC) simulations were performed. The hydrodynamic simulations were used to model the influence of the laser prepulses on the nanowire targets. The resulting expanded target density profile was then imported into 3D PIC simulations to describe the interaction with the main laser pulse. The simulation parameters were chosen to closely resemble the experimental conditions. The details of simulation parameters and methods are described in the supplementary material  \cite{SuppM}.

The angle-resolved proton energy spectra obtained from the PIC simulations are presented in Fig.~\ref{simulation_plots} (a) and (b) for the flat target (target-A) and the nanowire target (target-B), respectively. For both target configurations, proton acceleration occurs predominantly along the target normal, consistent with the target normal sheath acceleration  mechanism. The laser propagation direction is at $0^\circ$. At proton energies exceeding $10$~MeV, however, a pronounced difference between the two target types emerges: as shown in Fig. \ref{Fig:RCF image}(b) the nanowire targets produce a strongly forward-peaked proton emission, whereas the flat targets yield a substantially broader angular distribution, with the proton flux extending over angles from approximately $-30^\circ$ to $+60^\circ$ relative to the laser propagation direction. The proton energy spectra for flat and nanowire targets in the target normal direction integrated within the cone $\pm 10^\circ$ in Fig. \ref{simulation_plots} (c) again show an improvement in both cut-off energy and flux. The normalized proton angular flux obtained from the simulations is in good agreement with the RCF signal shown in Fig.~\ref{fig:raw RCF} and \ref{Fig:RCF image}. 

Most notably, the improved proton beam collimation observed experimentally from nanowire structures indicates that these structures provide an additional enhancement mechanism that has received relatively little attention in previous studies. Using PIC simulations, the increase in hot electron generation in the forward direction as reported in Parab \textit{et al.}  \cite{parab2026bright}, leads to enhancement of the toroidal magnetic field, $B_\theta$ at the rear surface. The toroidal magnetic field at the rear side of the target is shown in Figs. \ref{fig:B_field} (a)-(b), where the nanowire target in Fig. \ref{fig:B_field} (b) shows clear enhancement as compared to the flat target in Fig. \ref{fig:B_field} (a). The toroidal field enhancement in nanowire is approximately twice that for the flat target at $x=5.265\, \mathrm{\mu m}$, and $x=59.375\, \mathrm{\mu m}$ as shown in Fig. \ref{fig:B_field} (c) and (e), respectively.  The central region at $x=7.5\, \mathrm{\mu m}$ shows almost no enhancement as shown in Fig. \ref{fig:B_field} (d). This field deflects electrons over large transverse distances, after which they are attracted into the target by the strong sheath field and subsequently recirculate. In contrast, the proton beam accelerated by the sheath field away from the central region of the rear surface of the target is deflected outward, with only protons propagating close to the target normal axis being comparatively weakly affected. The deflection of higher energy protons are more pronounced in the nanowire target, as indicated by the $p_y$-$p_x$ phase-space distribution in Fig. \ref{fig:B_field} (g) as compared to Fig. \ref{fig:B_field} (f). The protons with momentum $p_y \sim 0.12 \, mc$ of Fig. \ref{fig:B_field} (g) and $p_y \sim 0.16 \, mc$ of Fig. \ref{fig:B_field} (h) that resemble the fountain are the protons deflected by the $e (\vec v \times \vec B)$ force. The net effect results in the production of a more collimated proton beam. 

The total energy of the forward propagating proton in Fig. \ref{simulation_plots} (a) and (b) is calculated to be $0.1 \, \mathrm{J}$ and $0.25 \, \mathrm{J}$, respectively. This corresponds to the conversion efficiency of $0.8 \, \%$ for flat targets and $2.1 \, \%$ for nanowire targets, exhibits $2.6\times$ enhancement in laser-to-proton conversion.

\section{Discussion}
For ultrahigh laser contrast ($10^{-13}$ at 50 ps time), achieved using the plasma mirror configuration, the experimentally measured proton spectra show a substantial enhancement in proton flux beyond 12 MeV for nanowire targets compared to flat foils [Figs. ~\ref{experimental_plots}(c) and \ref{fig:raw RCF}]. Quantitatively, the proton flux in the high-energy region (>12 MeV) is enhanced by a factor of $\sim 3.3$ for nanowires (B) relative to flat targets (A). This trend is reproduced and amplified in simulations, where the corresponding enhancement reaches nearly an order of magnitude [Fig. ~\ref{simulation_plots} (c)]. The agreement in qualitative behavior between experiment and simulation strongly suggests that the enhanced coupling of laser energy to hot electrons in nanostructured targets is the dominant mechanism driving the observed increase in high-energy proton production.

A comparison between high-contrast and lower-contrast ($10^{-13}$ vs $10^{-10}$ at 50 ps time) experimental data further highlights the critical role of target integrity during the main pulse interaction. As shown in Fig. ~\ref{experimental_plots} (b), when the contrast is reduced to in-built contrast of the laser ($10^{-10}$ at 50 ps) itself, the nanowire target still exhibits an enhancement (though reduced) in high-energy proton flux compared to the flat target irradiated under high-contrast conditions ($10^{-13}$ at 50 ps time). Even the femtosecond prepulses observed at 10 ps do not appear to have any effect. In any case,  they are close enough to the main peak, leaving insufficient time for hydrodynamic expansion that causes the preplasma (See Supplementary on the simulation). This observation suggests significant robustness of the nanowire structure on the rising edge of the pulse that continues to promote enhanced hot-electron generation and proton acceleration.

The shape of the angular spread of high energy protons [Fig. \ref{Fig:RCF image}] provides additional insight into the role of target structuring. The stronger directionality of higher energy (>12 MeV) proton beams with higher flux from nanowire targets (B) and good agreement with simulation is a major achievement of the experiment. The divergence half angle of 12.5 MeV protons for the flat target (our obtained value of $\sim17^\circ$) is very close to the results of Man\v{c}i\'{c} \textit{et al.}  \cite{mancic2010isochoric}.  Their targets are however 10 $\mu$m thick, much larger than ours. The  smaller angular divergence from our nanowire targets ($\sim 11^\circ$) clearly indicates the advantage of using  such structures for collimating the proton beams. 

Zeigler \textit{et al.}  \cite{zeil2010scaling} used a very thin flat foil with a two-pulse  excitation scheme  and  invoked  variations of the radiation pressure acceleration  process to reach  the highest energy value reported so far. On the other hand we are in the well established TNSA regime, but use structuring of the surface to boost the proton flux and directionality. Overall, the combined experimental and simulation results demonstrate that nanostructured targets particularly nanowire arrays, significantly enhance high-energy proton production and the energy cutoff under high-contrast laser irradiation. The enhancement arises not only from increased proton yield but also from improved beam directionality and modified spectral characteristics. These findings underline the importance of both laser contrast and target micro-structure in optimizing proton acceleration, and they highlight nanowire targets as a promising route toward efficient and controllable laser-driven ion sources.

\centerline{\large{\textsc{}}}
\centerline{\normalsize{\textsc{\uline{Acknowledgments}}}}

\noindent G.R.K. acknowledges major support for this research from Department of Atomic Energy through the grant “Physics and Astronomy (Project Identification No. RTI4002), Tata Institute of Fundamental Research” and partially from the Grant No. JBR/2021/00039 of the Anusandhan National Research Foundation (ANRF), both of the Government of India. {J.F.O and V.H appreciate access to the Karolina supercomputer at IT4Innovations (VŠB-TU, Czechia) through EuroHPC Joint Undertaking under project number EHPC-REG-2025R01-007 and Ministry of Education, Youth and Sports of the Czech Republic through the e-INFRA CZ (ID:90140) (OPEN-34-63, FTA-26-7, EU-25-87)}. {These works were partly supported by Contract No. PN23210105 funded by the Romanian Ministry of Research, Innovation and Digitalization and of the Extreme Light Infrastructure Nuclear Physics Phase II, a project co-financed by the Romanian Government and the European Union through the European Regional Development Fund and the Competitiveness Operational Program (Grant No. 1/07.07.2016, COP, ID 1334). Additionally, partial support was given by JSPS Core-to-Core Program, Grant Number JPJSCCA20230003. S.I and K.A.T acknowledge support from Nucleu Project (Grant No. PN 19060105) and IOSIN funds for Facilities of National Interest; and Project ELI-RO/DFG/2025-013 IATP-NP 2.0 funded by the Institute of Atomic Physics, Romania.} H.H acknowledges support from JSPS KAKENHI (grant number JP22H01205). We also acknowledge ELI-RO/RDI 16 DELPHI and ELI-RO/RDI 28 FLIGHT projects which supported part of the materials, equipment and manpower used in the experiments. We thank Mihail Cernaianu, Petru Ghenuche and Domenico Doria for encouragement and insights. Finally we thank the laser team and workshop team for their contribution in performing the experiment.

\centerline{\large{\textsc{}}}
\centerline{\normalsize{\textsc{\uline{Author Contributions}}}}
\centerline{\large{\textsc{}}}

\noindent S.D, S.I and J.F.O  contributed equally to this work. S.I
in collaboration with D.P and A.V, fabricated and characterized the nanowire targets. J.F.O and V.H performed the numerical simulations. A.P, S.R, Y.K participated in the experiment and
helped with the post processing. S.D did the data analysis in
collaboration with S.I, J.F.O and A.P. In the experimental campaign, L.T, D.S and S.I were
involved in the operation of the diagnostics and data acquisition. G.C, D.N, S.N, B.S, A.T, D.U lead the laser operation. R.P, K.S, H.H provided helpful
insights and advice. S.D, J.F.O, S.I and G.R.K wrote the
initial draft of the paper. All authors discussed the results,
contributed to the interpretation of the data, and reviewed the
manuscript. G.R.K, K.A.T and P.K.S conceived and supervised the entire project.

\bibliography{bibliography}

\end{document}


\title{SUPPLEMENTARY FOR -- ``Tens of MeV, collimated, bright fluxes of protons from ordered nano-structured targets in ultra-relativistic laser-matter interaction''}

\author{Sagar Dam}
\thanks{Email: \textcolor{blue}{sagar.dam@tifr.res.in}}
\affiliation{Tata Institute of Fundamental Research, Colaba, Mumbai 400005, India}

\author{Stefania Ionescu}
\thanks{Email: \textcolor{blue}{stefania.ionescu@eli-np.ro}}
\affiliation{ Extreme Light Infrastructure - Nuclear Physics (ELI-NP), ``Horia Hulubei'' National Institute for R\&D in Physics and Nuclear Engineering (IFIN-HH), 30 Reactorului Street, Bucharest-M\u{a}gurele, 077125, Romania}
\affiliation{National University of Science and Technology POLITEHNICA Bucharest, Splaiul Independentei no. 313, Bucharest, Romania}

\author{Jian Fuh Ong}
\thanks{Email: \textcolor{blue}{jianfuh.ong@eli-np.ro}}
\affiliation{ 
Extreme Light Infrastructure - Nuclear Physics (ELI-NP), ``Horia Hulubei'' National Institute for R\&D in Physics and Nuclear Engineering (IFIN-HH), 30 Reactorului Street, Bucharest-M\u{a}gurele, 077125, Romania}

\author{Ameya Parab}
\thanks{Email: \textcolor{blue}{ameya.parab@tifr.res.in}}
\affiliation{Tata Institute of Fundamental Research, Colaba, Mumbai 400005, India}

\author{Sk Rakeeb}
\affiliation{Tata Institute of Fundamental Research, Colaba, Mumbai 400005, India}

\author{Hideaki Habara}
\affiliation{University of Osaka, Yamadaoka, Suita, Osaka 565-0871, Japan}

\author{Gabriel Cojocaru}
\affiliation{ 
Extreme Light Infrastructure - Nuclear Physics (ELI-NP), ``Horia Hulubei'' National Institute for R\&D in Physics and Nuclear Engineering (IFIN-HH), 30 Reactorului Street, Bucharest-M\u{a}gurele, 077125, Romania}
\affiliation{National Institute for Laser, Plasma and Radiation Physics, CETAL-PW Department, 409 Atomistilor Street, Magurele, Romania, 077125}

\author{Vojtěch Horný}
\affiliation{ 
Extreme Light Infrastructure - Nuclear Physics (ELI-NP), ``Horia Hulubei'' National Institute for R\&D in Physics and Nuclear Engineering (IFIN-HH), 30 Reactorului Street, Bucharest-M\u{a}gurele, 077125, Romania}

\affiliation{Faculty of Nuclear Sciences and Physical Engineering, Czech Technical University in Prague, Břehová 7, 115 19 Prague 1, Czechia}

\author{Dmitrii Nistor}
\affiliation{ 
Extreme Light Infrastructure - Nuclear Physics (ELI-NP), ``Horia Hulubei'' National Institute for R\&D in Physics and Nuclear Engineering (IFIN-HH), 30 Reactorului Street, Bucharest-M\u{a}gurele, 077125, Romania}
\affiliation{National University of Science and Technology POLITEHNICA Bucharest, Splaiul Independentei no. 313, Bucharest, Romania}

\author{Saidbek Norbaev}
\affiliation{ 
Extreme Light Infrastructure - Nuclear Physics (ELI-NP), ``Horia Hulubei'' National Institute for R\&D in Physics and Nuclear Engineering (IFIN-HH), 30 Reactorului Street, Bucharest-M\u{a}gurele, 077125, Romania}

\author{Rudrajyoti Palit}
 \affiliation{Tata Institute of Fundamental Research, Colaba, Mumbai 400005, India}

\author{Daniel Popa}
\affiliation{ 
Extreme Light Infrastructure - Nuclear Physics (ELI-NP), ``Horia Hulubei'' National Institute for R\&D in Physics and Nuclear Engineering (IFIN-HH), 30 Reactorului Street, Bucharest-M\u{a}gurele, 077125, Romania}

\author{Deepak Sangwan}
\affiliation{ 
Extreme Light Infrastructure - Nuclear Physics (ELI-NP), ``Horia Hulubei'' National Institute for R\&D in Physics and Nuclear Engineering (IFIN-HH), 30 Reactorului Street, Bucharest-M\u{a}gurele, 077125, Romania}

\author{Klaus Spohr}
\affiliation{ 
Extreme Light Infrastructure - Nuclear Physics (ELI-NP), ``Horia Hulubei'' National Institute for R\&D in Physics and Nuclear Engineering (IFIN-HH), 30 Reactorului Street, Bucharest-M\u{a}gurele, 077125, Romania}

\author{Bianca Stan}
\affiliation{ 
Extreme Light Infrastructure - Nuclear Physics (ELI-NP), ``Horia Hulubei'' National Institute for R\&D in Physics and Nuclear Engineering (IFIN-HH), 30 Reactorului Street, Bucharest-M\u{a}gurele, 077125, Romania}

\author{Antonia Toma}
\affiliation{ 
Extreme Light Infrastructure - Nuclear Physics (ELI-NP), ``Horia Hulubei'' National Institute for R\&D in Physics and Nuclear Engineering (IFIN-HH), 30 Reactorului Street, Bucharest-M\u{a}gurele, 077125, Romania}

\author{Lucian Tudor}
\affiliation{ 
Extreme Light Infrastructure - Nuclear Physics (ELI-NP), ``Horia Hulubei'' National Institute for R\&D in Physics and Nuclear Engineering (IFIN-HH), 30 Reactorului Street, Bucharest-M\u{a}gurele, 077125, Romania}

\author{Daniel Ursescu}
\affiliation{ 
Extreme Light Infrastructure - Nuclear Physics (ELI-NP), ``Horia Hulubei'' National Institute for R\&D in Physics and Nuclear Engineering (IFIN-HH), 30 Reactorului Street, Bucharest-M\u{a}gurele, 077125, Romania}

\author{Adrian Vatcu}
\affiliation{ 
Extreme Light Infrastructure - Nuclear Physics (ELI-NP), ``Horia Hulubei'' National Institute for R\&D in Physics and Nuclear Engineering (IFIN-HH), 30 Reactorului Street, Bucharest-M\u{a}gurele, 077125, Romania}

\author{Keita Yamanaka}
\affiliation{University of Osaka, Yamadaoka, Suita, Osaka 565-0871, Japan}

\author{Prashant Kumar Singh}
\affiliation{Tata Institute of Fundamental Research Hyderabad, 36/P, Gopanpally Village, Serilingampally Mandal, Hyderabad, Telangana 500046, India}

\author{Kazuo.A.Tanaka}
\affiliation{ 
Extreme Light Infrastructure - Nuclear Physics (ELI-NP), ``Horia Hulubei'' National Institute for R\&D in Physics and Nuclear Engineering (IFIN-HH), 30 Reactorului Street, Bucharest-M\u{a}gurele, 077125, Romania}
\affiliation{University of Osaka, Yamadaoka, Suita, Osaka 565-0871, Japan}

\author{G.Ravindra Kumar}
    \thanks{Email: \textcolor{blue}{grk@tifr.res.in}}
    \thanks{(corresponding author)}
    \affiliation{Tata Institute of Fundamental Research, Colaba, Mumbai 400005, India}


\date{\today}

\maketitle

\section{Working principle of a Thomson parabola}

In a Thomson parabola (TP) spectrometer, accelerated ions are deflected by combined electric and magnetic fields according to their kinetic energy and charge-to-mass ratio, resulting in characteristic parabolic traces on the image plate (IP) detector. The relation between deflections due to the magnetic and electric fields ($\Delta x_B \ \&\ \Delta x_E$ respectively) is given by
\begin{equation}
\frac{\Delta x_B^2}{\Delta x_E}
= k\left(\frac{B^2}{E}\right)\left(\frac{q}{m}\right),
\end{equation}
where \(B\) and \(E\) denote the magnetic and electric field strengths, respectively, \(q/m\) is the ion charge-to-mass ratio, and \(k\) is a geometry-dependent constant determined by the TP configuration.

Lower-energy ions experience larger deflections and therefore strike the detector farther from the zero-deflection point, whereas higher-energy ions undergo smaller deflections and follow trajectories closer to a straight path. The curvature of the parabolic trace is governed by the \(q/m\) ratio of the ion species, with ions possessing a higher \(q/m\) exhibiting a reduced curvature.

\begin{figure}
\centering
\includegraphics[width=\columnwidth]{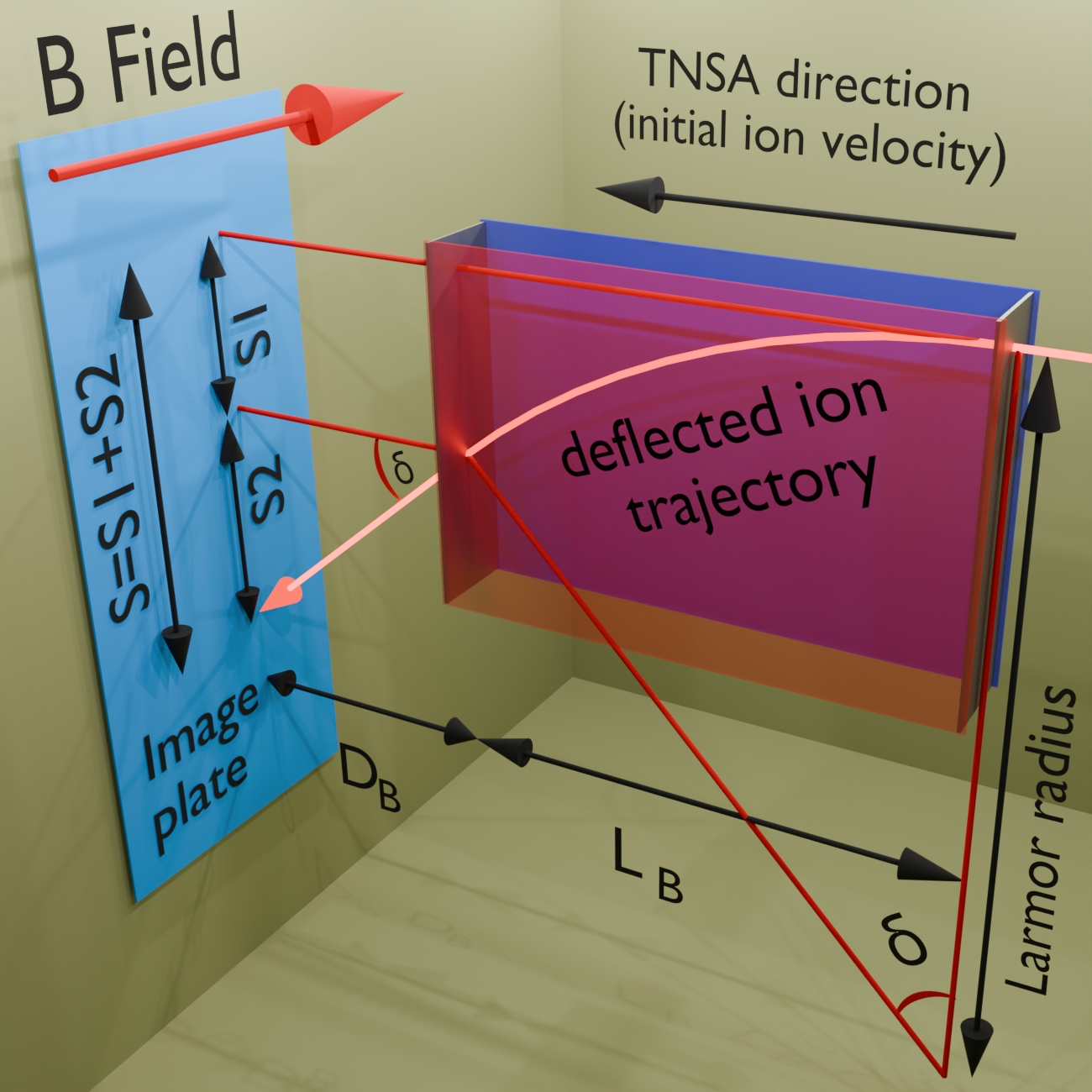}
\caption{Deflection of ions due to the magnetic field in the Thomson parabola spectrometer.}
\label{fig:proton path}
\end{figure}

The ion kinetic energy is determined from the total magnetic deflection distance \(S_{\mathrm{total}}\), measured from the zero-deflection point on the image plate (or other detector), as illustrated in Fig.~\ref{fig:proton path}. The total deflection is expressed as
\begin{equation}
S_{\mathrm{total}} = S_1 + S_2,
\end{equation}
where
\begin{equation}
S_1 = R_L - \sqrt{R_L^2 - L_B^2},
\end{equation}
and
\begin{equation}
S_2 = D_B \tan \delta.
\end{equation}
Here, \(L_B\) is the magnetic field length, \(D_B\) is the distance between the exit edge of the magnet and the image plate (IP) detector, \(R_L\) is the relativistic Larmor radius [eq:\ref{Larmor radius}] corresponding to an ion of kinetic energy \(E\) and $\delta$ is the deflection angle [eq:\ref{deflection angle}].

\noindent The relativistic Larmor radius is given by~ \cite{tanaka2005IPcalibration}
\begin{equation}
R_L(E) =
\frac{m_0 c}{q B}
\sqrt{
\left(
\frac{E + m_0 c^2}{m_0 c^2}
\right)^2 - 1
},
\label{Larmor radius}
\end{equation}
where \(m_0\) is the rest mass of the ion, \(q\) is the ion charge, and \(c\) is the speed of light. The deflection angle \(\delta\) is defined as
\begin{equation}
\delta = \tan^{-1}\left(\frac{L_B}{\sqrt{R_L^2 - L_B^2}}\right),\ \ (\text{When }L_B < R_L)
\label{deflection angle}
\end{equation}

\section{Analyzing the image plate (IP)}

\begin{figure}[h]
    \centering
    \includegraphics[width=\linewidth]{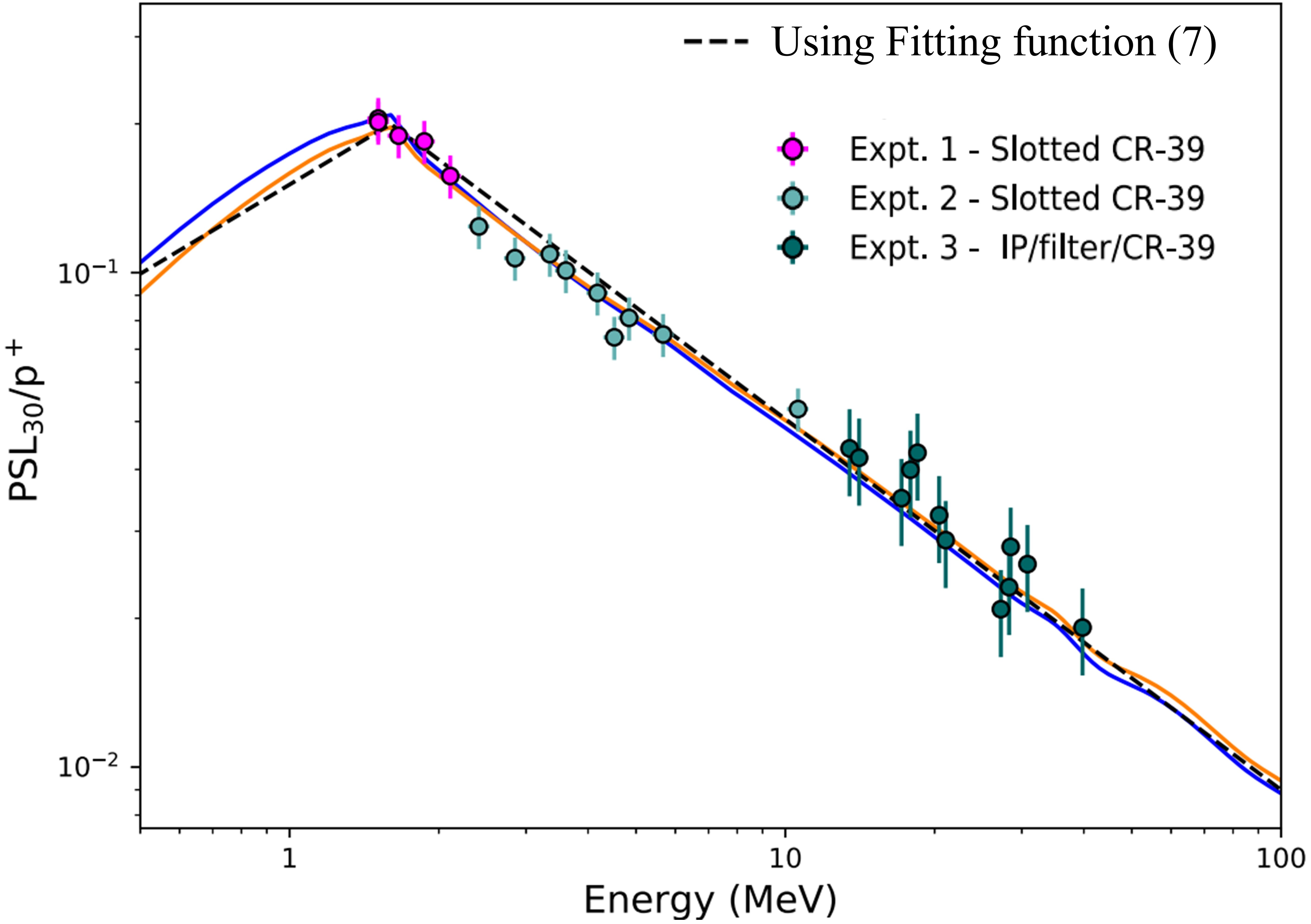}
    \caption{TR-IP calibration for different proton energy and fitting (black dashed line) the response function (eq \ref{IP calibration equation}) from Ref: Martin \textit{et al}. \cite{Calibration_of_TR_IP_for_ions}}
    \label{fig:IP calibration for proton}
\end{figure}

The photo-stimulated luminescence (PSL) signal recorded on the IP depends on the ion energy and must be corrected to accurately reconstruct the ion energy spectrum. Previous calibration studies by Martin \textit{et al.}~ \cite{Calibration_of_TR_IP_for_ions} [see plot in fig:\ref{fig:IP calibration for proton} from their study] for the same TR-type IP provide the proton response function as
\begin{equation}
\mathrm{PSL}_{30}^{\mathrm{proton}} =
\begin{cases}
0.151\, E_p^{0.6}, & E_p < 1.6~\mathrm{MeV}, \\
0.284\, E_p^{-0.75}, & E_p \ge 1.6~\mathrm{MeV},
\end{cases}
\label{IP calibration equation}
\end{equation}
where \(\mathrm{PSL}_{30}^{\mathrm{proton}}\) denotes the PSL signal measured 30 minutes after irradiation and \(E_p\) is the proton energy in MeV.

Due to spontaneous fading of the IP signal over time, the PSL measured at an arbitrary time \(t\) (in minutes) after exposure is related to \(\mathrm{PSL}_{30}^{\mathrm{proton}}\) as~ \cite{time_response_of_TR_IP}
\begin{equation}
\mathrm{PSL}^{\mathrm{proton}}(t)
=
\left(
\frac{t}{30}
\right)^{-0.161}
\mathrm{PSL}_{30}^{\mathrm{proton}}.
\end{equation}

By incorporating the IP response function, temporal fading correction, and the acceptance solid angle \(d\Omega\) (defined by the \(200~\mu\mathrm{m}\) pinhole placed before the TP in our experiment), the differential proton yield along the target normal direction ($\hat n$) within the energy interval \([E, E + dE]\) is obtained as
\begin{equation}
S(E,\hat n)
=
\frac{
\mathrm{sig}_{[E,E+dE]}
\times \mathrm{fading~factor}
\times \mathrm{PSL}(t)
}{
d\Omega
},
\end{equation}
where \(\mathrm{sig}_{[E,E+dE]}\) corresponds to the IP signal associated with magnetic deflection between \(l_p\) and \(l_p + dl_p\).

The integrated proton spectrum, representing the cumulative number of protons with energies below \(E\), is then given by
\begin{equation}
I(E,\hat n)
=
\int_{E_{\mathrm{low}}}^{E}
S(\epsilon,\hat n)\, d\epsilon.
\end{equation}

(Where $E_\text{low}$ is the lower limit of the spectrometer detection range.)

\section{Working principle of Radio-Chromic Film (RCF) and energy calibration for stack}

\begin{table}[h]
    \centering
    \renewcommand{\arraystretch}{1.4}
    \footnotesize
    \begin{tabular}{|c |c |c|}
        \hline
        \hline
        \textbf{Layer} & \textbf{Thickness (mm)} & \textbf{Energy (MeV)} \\
        \hline
        \hline
        Aluminium & 0.012 & -- \\
        \hline
        HD-V2 (1) & 0.109 & 1.1 \\
        \hline
        HD-V2 (2) & 0.109 & 3.2 \\
        \hline
        HD-V2 (3) & 0.109 & 4.57 \\
        \hline
        HD-V2 (4) & 0.109 & 5.67 \\
        \hline
        PEHD & 0.136 & -- \\
        \hline
        HD-V2 (5) & 0.109 & 7.5 \\
        \hline
        2-PEHD & 0.272 & -- \\
        \hline
        HD-V2 (6) & 0.109 & 9.66 \\
        \hline
        Cu (1) & 0.1 & -- \\
        \hline
        HD-V2 (7) & 0.109 & 12.3 \\
        \hline
        Cu (2) & 0.1 & -- \\
        \hline
        HD-V2 (8) & 0.109 & 14.51 \\
        \hline
        Cu (3) & 0.1 & -- \\
        \hline
        EBT-3 (9) & 0.26 & 16.75 \\
        \hline
        Cu (4) & 0.1 & -- \\
        \hline
        EBT-3 (10) & 0.26 & 18.65 \\
        \hline
        Cu (5) & 0.3 & -- \\
        \hline
        EBT-3 (11) & 0.26 & 23.2 \\
        \hline
        Cu (6) & 0.3 & -- \\
        \hline
        EBT-3 (12) & 0.26 & 27.55 \\
        \hline
        Cu (7) & 0.3 & -- \\
        \hline
        EBT-3 (13) & 0.26 & 31.4 \\
        \hline
        Cu (8) & 0.3 & -- \\
        \hline
        EBT-3 (14) & 0.26 & 35 \\
        \hline
        \hline
    \end{tabular}
    \caption{Used layer structure of the RCF stack and corresponding proton energies.}
    \label{tab:RCF_stacks}
\end{table}

Radiochromic films (RCFs) are self-developing dosimetric detectors widely used for spatially and spectrally resolved measurements of energetic ions in laser–plasma interaction experiments. When exposed to ionizing radiation, RCFs undergo a radiation-induced polymerization process in the active layer, leading to a permanent change in optical density (OD) that is proportional to the deposited dose. This response does not require chemical processing and is stable over time, making RCFs particularly suitable for high-repetition-rate and high-fluence environments.

The dose deposited in each RCF layer depends on the stopping power of the incident ions and the depth of penetration in the film material. By stacking multiple RCF layers with known thicknesses and material composition, energy-resolved information of the ion beam can be obtained, as ions of different energies deposit their Bragg peak at different depths within the stack. Consequently, each RCF layer effectively corresponds to a narrow energy window, determined by ion stopping calculations.

The exposed RCF(s) were digitized using a high-resolution flatbed image scanner under identical scanning conditions for all shots to ensure consistency and reproducibility. In the present analysis, the scanned images were used to extract relative signal intensities, which were subsequently normalized to enable comparison of spatial and angular features of the proton emission. No absolute calibration of optical density to absorbed dose or proton flux was performed. Instead, the RCF data were used to qualitatively and comparatively assess proton beam  profiles and relative yield variations at selected energy layers within the RCF stack.

In this work, RCF stacks were used to complement the Thomson parabola measurements by providing two-dimensional spatial and angular profiles of the accelerated proton beam. The corresponding calibration chart used to convert RCF stacking position to proton energy is presented in Table:\ref{tab:RCF_stacks}.

\section{Simulation details}

\begin{table}
\centering
\footnotesize
\begin{ruledtabular}
\begin{tabular}{ll}
\multicolumn{2}{c}{\textsc{Simulation parameters and values}} \\
\hline
\multicolumn{2}{c}{\textit{\textcolor{violet}{\textbf{\uline{Geometry}}}}} \\[4pt]
\textsc{Flash 4.8} & $\begin{cases} \text{Simulation box size: } 15 \times 15 \times 15 \ \mathrm{\mu m}^3 \\ \text{Min. Grid res: } \left\{^{\text{Flat: } 58.6 \ \times \ 29.3 \ \times \ 58.6 \ \text{nm}^3}_{\text{NW: } 29.3 \ \times 29.3 \times 29.3 \ \text{nm}^3} \right. \\ \text{Flux limiter: 0.08} \end{cases}$ \\

 & \\

\textsc{Picongpu 0.8.0} & $\begin{cases}\text{Grid resolution:}\ \left\{^{\text{Flat: }29\times14.6\times29\text{ nm}^3}_{\text{NW: }14.6\times14.6\times14.6\text{ nm}^3}\right. \\ \text{Temporal resolution: } 5.28 \times 10^{-18} \,  \text{s} \\ \text{No of particle-per-cell}:\ 8 \\ \end{cases}$ \\ 

\hline
\multicolumn{2}{c}{\textit{\textcolor{violet}{\textbf{\uline{Laser}}}}} \\[4pt]
Wavelength & $\lambda_L: 800~\mathrm{nm}$\\
Intensity & $I_0 = 2.8 \times 10^{21}~\mathrm{W/cm^2};\ \ a_0=36$\\
$\theta_\text{incidence}$ & $15^\circ\ \ \ \ \ \ \ \ \ \ \ \ \ \ \ \ \ \ \ \ \ \ \ \ \ \ \ \ \ \ \ \ \ \ \ \ \ \ \ \ \ \ \ \ \ \ \ \ \ $\\
Pulse duration & $\tau_L$ = 25 fs\\
Focal spot & $d_L$ = 4.2 $\mu$m\\
Polarization & $p-$polarized \\
\hline
\multicolumn{2}{c}{\textit{\textcolor{violet}{\textbf{\uline{General parameters}}}}} \\[4pt]
Critical density & $n_c(\omega_L) = 1.74 \times 10^{21}~\mathrm{cm^{-3}}$ \\
Ion density & $n_i = 9.13 \times 10^{22}~\mathrm{cm^{-3}}$ \\
Electron density & $n_e = 2.56 \times 10^{24}~\mathrm{cm^{-3}}$ \\
\hline
\multicolumn{2}{c}{\textit{\textcolor{violet}{\textbf{\uline{Target dimensions}}}}} \\[4pt]
Flat target & Thickness: 2 $\mu$m \\[8pt]
NW \hyperref[type 1 NT NW]{type 1}
 & $\begin{cases} \text{substrate} = 1.2\ \mu\text{m}\\ \text{NW length}= 1\ \mu\text{m} \\ \text{NW diameter}= 0.1\ \mu\text{m} \\\text{Separation}= 0.3\ \mu\text{m} \\ \text{Array size}=30\times30\end{cases}$ \\
\hline
\multicolumn{2}{c}{\textit{\textcolor{violet}{\textbf{\uline{Hydrocarbon contamination}}}}} \\[4pt]
Hydrogen density & $n_\mathrm{H} = 27 n_\mathrm{cr}$ \\
Carbon density & $n_\mathrm{C} = 159 n_\mathrm{cr}$ \\
Layer thickness & $d_\text{contamination} = 50$ nm 
\end{tabular}
\caption{\liber{Simulation Parameters\\
}}
\label{tab:sim_params}
\end{ruledtabular}
\end{table}

We performed combined three-dimensional radiation hydrodynamic (3D RHD) and Particle-In-Cell (3D PIC) to model the experiments. 3D RHD simulations were performed using \textsc{flash} code version 4.8  \cite{flash}. We use Adaptive Mesh Refinement (AMR) with 6 levels of refinement on the mass density, achieving a minimum resolution of $29 \, \mathrm{nm}$. The pressure and opacity of nickel were obtained from tabulated equation of state \textsc{sesame} and \textsc{tops} opacity data. The ionization state was calculated using the Thomas-Fermi model. We used the Lee-More model for heat exchange and electron heat conduction, with a flux limiter of $0.08$. The laser propagation is modeled using Ray Tracing, and the energy is dopisted via inverse Bremsstrahlung. \textsc{flash} code does not incorporate multi-photon or barrier suppression ionization, therefore, the short femtosecond prepulse would only heat the target via inverse Bremsstrahlung. Due to its short duration, it contributes very little to the plasma expansion.

For the flat target, the ablation starts from $t \sim -50 \, \mathrm{ps}$ and from $t \sim -311 \, \mathrm{ps}$ for the nanowire target. The onset of ablation is determined by comparing the absorbed fluence $F_\mathrm{abs}(t) = \int_0^{t} A(t') I(t') dt'$ to the fluence threshold $F_\mathrm{th}^\mathrm{long} \sim (\kappa t_p)^{1/2} \epsilon_b n_a / A$ in the long pulse ablation regime along the contrast profile  \cite{Gamaly2002,Gamaly2004}. Here, $A(t)$ is the time dependent absorption coefficient, $\kappa$ is the thermal diffusion coefficient, $\epsilon_b$ is the inter-atomic binding energy, $n_a$ is the number density of the neutral ion and $A$ is the plasma absorption coefficient. The pulse duration $t_p$ here is equivalent to the irradiation duration from the beginning of the contrast profile. Once the accumulated fluence exceeds the ablation threshold, i.e., $F_\mathrm{abs} > F_\mathrm{th}^\mathrm{long}$ the ablation time is determined.

\begin{figure}
    \centering
    \includegraphics[width=0.75\linewidth]{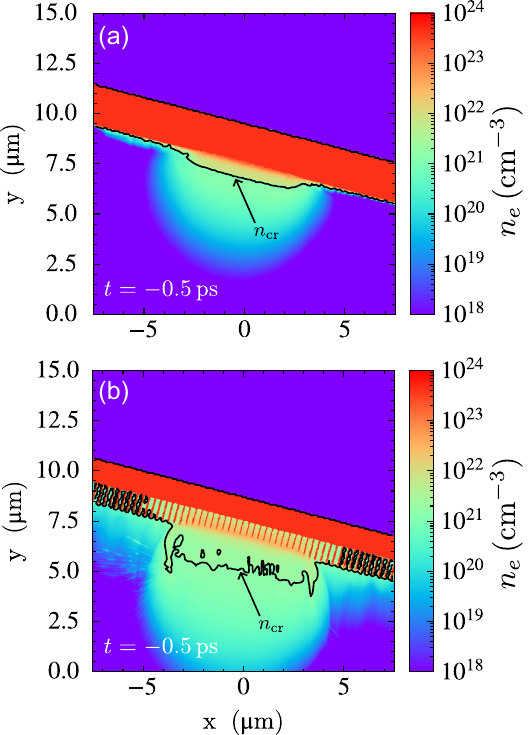}
    \caption{Snapshots of electron density of 3D RHD simulation at $t = -0.5 \, \mathrm{ps}$ for (a) flat target, and (b) nanowire target. The solid line indicates the critical density $n_\mathrm{cr}=1.74 \times 10^{21} \, \mathrm{cm^{-3}}$ at $\lambda_L = 800 \, \mathrm{nm}$. }
    \label{fig:FIGSX}
\end{figure}

The flat nickel target has a thickness of $2 \, \mathrm{\mu m}$. The nickel nanowire is an aligned $30 \times 30$ array with diameter $d = 100\, \mathrm{nm}$, length $L = 1 \,\mathrm{\mu m}$, and center-to-center separation $300 \, \mathrm{nm}$ that attaches to the $1.2 \,\mathrm{\mu m}$ substrate of the same material. The laser is incident at $15^\circ$ from the target normal. The simulation ends at $t \sim -0.24 \, \mathrm{ps}$, where $I_0[\mathrm{W \, cm^{-2}}] \lambda^2_\mathrm{\mu m} \simeq 10^{17}$. The electron density profile at $t = -0.5 \, \mathrm{ps}$ for the flat target is shown in Fig. \ref{fig:FIGSX} (a) and the nanowire target in Fig. \ref{fig:FIGSX} (b). The mass density profile is then converted to PIC input using \textsc{flash2openpmd}  \cite{flash2openPMD}. The grid resolution is reduced through cascading interpolation up to six levels of refinement.

The 3D PIC simulation is carried out using the \textsc{picongpu} code version 0.8.0  \cite{PIConGPU}. The simulation of the flat target has a resolution of $29 \times 14.6 \times 29 \, \mathrm{nm^3}$, while for the nanowire is $14.6 \times 14.6 \times 14.6 \, \mathrm{nm^3}$. Each macro-particle species has 8 particle-per-cell, and the shape function used is piecewise quartic spline. To mitigate numerical heating, binomial current interpolation is employed. The target temperatures and charge states obtained from the hydrodynamic simulation vary and are non-uniformly distributed. For simplicity, we assume a uniform temperature and a fully ionized target. The hydrocarbon contamination on the target rear side is assumed to have thickness of $50 \, \mathrm{nm}$ with density $n_\mathrm{C} = 159 n_\mathrm{cr}$, and $n_\mathrm{H} = 27 n_\mathrm{cr}$. We note that the profile of the contamination layer and the density would affect the proton energy spectrum. We assumed that the contamination has a uniform profile at the moment.

The temporal profile of the main pulse includes the rising edge of the Gaussian pulse from $t \gtrsim -0.24 \, \mathrm{ps}$. The pulse duration of the main laser pulse is $t_L = 25 \, \mathrm{fs}$ with a FWHM focal spot diameter of $d_\mathrm{FWHM} = 4.2 \, \mathrm{\mu m}$, and wavelength of $\lambda_L = 800 \, \mathrm{nm}$. The laser is p-polarized and focused on the target surface with a peak intensity of $I_0 = 2.8 \times 10^{21} \, \mathrm{W \, cm^{-2}}$, which corresponds to the normalized amplitude of $a_0 = 36$. Table \ref{tab:sim_params} summarizes the simulation parameters used.

\bibliography{bibliography}